\documentclass[twocolumn,superscriptaddress,amssymb,amsmath,nobibnotes,aps,prd,showpacs,nofootinbib]{revtex4}%
\pdfoutput=1

\usepackage{graphicx}
\usepackage{epsf}
\usepackage{bm}
\usepackage{amsmath,hyperref}
\usepackage{amsfonts}
\usepackage{amssymb}
\usepackage{epstopdf}
\usepackage{natbib}%
\usepackage{rotating}
\usepackage{pdflscape}
\usepackage{adjustbox}
\usepackage[toc,page]{appendix}
\providecommand{\U}[1]{\protect\rule{.1in}{.1in}}
\newcommand{\be}{\begin{equation}}
\newcommand{\ee}{\end{equation}}

\newcommand{\mincir}{\raise
-3.truept\hbox{\rlap{\hbox{$\sim$}}\raise4.truept\hbox{$<$}\ }}
\newcommand{\magcir}{\raise
-3.truept\hbox{\rlap{\hbox{$\sim$}}\raise4.truept\hbox{$>$}\ }}

\usepackage{color}

\begin{document}

\title{\textcolor{blue}{\small{$\quad$\hspace{1cm}Eur. Phys. J. C \textbf{77} (2017) 734 }}\newline$\;$\newline Observational constraints on dark matter-dark energy scattering cross section}
\author{Suresh Kumar}
\email{suresh.kumar@pilani.bits-pilani.ac.in}
\affiliation{Department of Mathematics, BITS Pilani, Pilani Campus, Rajasthan-333031, India}

\author{Rafael C. Nunes}
\email{rafadcnunes@gmail.com}
\affiliation{Departamento de F\'isica, Universidade Federal de Juiz de Fora, 36036-330, Juiz de Fora, MG, Brazil}


\begin{abstract}
In this letter, we report precise and robust observational constraints on dark matter-dark energy scattering cross section, using
the latest data from cosmic microwave background (CMB) Planck temperature and polarization,
baryon acoustic oscillations (BAO) measurements and weak gravitational lensing data from Canada-France-Hawaii Telescope Lensing Survey
(CFHTLenS). The scattering scenario consists of a pure momentum exchange between the dark components, and 
we find $\sigma_d < 10^{-29} \, {\rm cm^2}$ at 95\% CL from the joint analysis (CMB + BAO + CFHTLenS), 
for typical dark matter particle mass of the order 10 ${\rm GeV}/c^2$. We notice that the scattering among the dark components
may influence the growth of large scale structure in the Universe, leaving the background cosmology unaltered.

\end{abstract}

\maketitle

\section{Introduction}
\label{sec:intro}

Cosmological observations reveal that approximately 95\% of the energy content of our Universe is unknown, and this matter/energy content 
is usually termed the dark sector of the Universe.
Constraints from Planck team \cite{Planck2015} show that 25\% of this content is in the form of a non-relativistic matter called dark matter (DM).
The other part of this dark sector is in the form of an exotic component, dubbed dark energy (DE), 
with negative pressure responsible for accelerated expansion of the Universe at late times \cite{DE1,DE2}. 
The particle physics experiments are yet to discover suitable particle candidates for the two dark components, 
and it is one of the greatest challenges in the contemporary physics.
The popular candidate of DE is a positive cosmological constant \cite{cc1,cc2,cc3} that 
suffers from theoretical inconsistencies, and  consequently several alternative models have been proposed in the literature 
to explain the late time accelerated expansion of the Universe \cite{Copeland06}. \\ 

Motivated to solve/assuage the theoretical problems in the $\Lambda$CDM model, 
models are proposed/studied in the literature where the dark components do not evolve separately but interact non-gravitationally with one another 
throughout the expansion of the Universe (see \cite{DE_DM_1,DE_DM_2} for reviews).
Essentially these interacting models invoke energy exchange between the dark components and a small
degree of momentum exchange. These scenarios have been intensively investigated in the literature for several dark sector coupling functions 
and observational approximations \cite{coupled01,coupled02,coupled03,coupled04,coupled05,coupled06,coupled07,coupled08,coupled09,coupled10,coupled11,
coupled12,coupled13,coupled14,coupled15,coupled16,coupled17}. 
It has recently been shown that the current observational data can favor the late time interaction between 
DE and DM \cite{Salvatelli:2014zta,Sola:2015,Murgia:2016ccp,Pigozzo,eu1,eu2,Valentino,Ferreira}.

Models with energy exchange modify the cosmology at background level (expansion history) 
and perturbative level (growth of structures). A scenario with a pure momentum exchange 
between the dark components is presented in \cite{Simpson}, where the
energy exchange between the dark components is negligible. Thus, the expansion history is the same as for a non-interacting scenario.
This class of interaction considers an elastic scattering of the DM particles
with DE. For, the DE presents a low density ($\sim 10^{-47}\; \text{GeV}^{4}$), and DM particles exhibit 
nonrelativistic velocities dispersion, that is, DM is cold, or at least a considerable part of the DM is cold. Therefore, an elastic scattering appears as a natural modeling of the dark sector physics.   
On the other hand, elastic scattering is the most abundant process at the energy scales of interest.
The model is independent of the microphysics involved in the scattering process, or the DE nature.
Consequences of the elastic scattering interaction between the dark components on the growth of large scale structure are presented in \cite{Simpson} while 
its effects on linear and nonlinear structure formation via N-body simulations are investigated in \cite{Baldi1,Baldi2}. 
In the next section, we present more details about this interaction model.

The aim of this paper is to constrain the DM-DE scattering cross section (as well as the equation of state parameter of 
DE in this context) with precision and robustness using the latest observational data from CMB Planck temperature and polarization,
BAO measurements and CFHTLenS data. In next section, we introduce the model with the elastic scattering between DM and DE.
We present the results and the related discussion of our analysis in Section \ref{results}. 
In the final section, we summarize the findings of our study with future perspectives.

\section{Dark matter-dark energy scattering models}
The interaction between particles can be quantified by the cross section (the likelihood within
an area transverse to their relative motions) within which the particles must meet in order to scatter from each other,
implying a transfer of momentum and/or energy, or lead to the creation of new particles. 
At low energies the cross section can be well described by a process of elastic scattering. For instance, the photon-baryons fluid
(essentially photon-electrons) in the early Universe is given in terms of the Thomson cross section, $\sigma_T \simeq 10^{-25}$ cm$^2$. 
In general, the drag force can be written as

\begin{align}
\label{drag_force}
\textbf{F} = -(1+w) \sigma \gamma^2 \rho \textbf{v}, 
\end{align}
where $v$ is the velocity of the particle traversing the fluid, and $\gamma$ is the Lorentz factor. For instance,
the photon-electron coupling ($w = 1/3$) is given by $\textbf{F} = - 4/3 \sigma_T v \rho_{\gamma}$. 

Considering the drag force \ref{drag_force} as the only non-gravitational force between DE and DM,
the linear perturbation theory equations (in conformal newtonian gauge) for DE-DM interaction fluids 
are given by \cite{Simpson,Ma_Bertschinger}

\begin{align}
\label{theta1}
\theta'_{\rm de} = 2 \mathcal{H} \theta_{\rm de} + k^2 \frac{\delta_{\rm de}}{1+w_{\rm de}} + k^2 \Psi \nonumber \\ 
-a n_{\rm dm} \sigma_{d}(\theta_{\rm de}-\theta_{\rm dm}),
\end{align}
and 

\begin{align}
\label{theta1.1}
\theta'_{\rm dm}  = - \mathcal{H} \theta_{\rm dm} + k^2 \Psi + \nonumber \\  
\frac{\rho_{\rm de}}{\rho_{\rm dm}} (1+w_{\rm de}) a n_{\rm dm} \sigma_{ d}(\theta_{\rm de}-\theta_{\rm dm}),
\end{align}
where $n_{\rm dm}$ is the proper number density of DM particles, $\sigma_{d}$ is the scattering cross section 
between DE and DM. 

The velocity perturbation above exhibits a new drag term 

\begin{align}
\label{drag}
S = a n_{\rm dm} \sigma_{d}(\theta_{\rm de}-\theta_{\rm dm}).
\end{align}

It represents the DE fraction which is subject to scattering per unit time. 
In this scenario, the equation of continuity, $\delta_{i=\rm dm,de}=\delta \rho_{i=\rm dm,de}/\rho_{i=\rm dm,de}$, 
remains unchanged and follows the standard evolution. As already commented above, this scenario assumes 
that the dark components are not coupled at background level. Here, the interaction in the dark sector is quantified by the 
drag term $S$, and only acts via the velocity perturbation equations.

Once that $\rho_{\rm dm} = m_{\rm dm} n_{\rm dm}$ for DM particles, where
$m_{\rm dm}$ stands for the mass of a typical DM particle. Then, eq. (\ref{drag}) can be rewritten as 
 
\begin{align}
S = a \rho_{\rm dm} \xi (\theta_{\rm de}-\theta_{\rm dm}),
\end{align}
where we have defined
 
\begin{align}
\label{xi}
\xi =  \frac{\sigma_d}{m_{\rm dm}},
\end{align}
with dimensions/units of barn $\cdot$ $c^2$/GeV, as the characteristic parameter of the drag term. 
 
\section{Results}
\label{results}
 
In order to constrain the cross section of elastic scattering between DE and DM, we consider the following data sets (briefly described) sensitive 
at the perturbations level.\\

\noindent\textbf{CMB}: We use the full Planck 2015 data \cite{Planck2015} comprised of temperature (TT), 
polarization (EE) and the cross correlation of temperature and polarization (TE) together with the CMB lensing power spectrum. 
\\

\noindent\textbf{BAO}: We use the BAO measurements from the  Six  Degree  Field  Galaxy  Survey  (6dF) \cite{bao1}, 
the  Main  Galaxy  Sample  of  Data  Release 7  of  Sloan  Digital  Sky  Survey  (SDSS-MGS) \cite{bao2}, 
the  LOWZ  and  CMASS  galaxy  samples  of  the Baryon  Oscillation  Spectroscopic  Survey  (BOSS-LOWZ  and  BOSS-CMASS,  
respectively) \cite{bao3},  and the distribution of the LymanForest in BOSS (BOSS-Ly) \cite{bao4}.
These data points are summarized in Table I of \cite{eu3}. 
\\

\noindent\textbf{CFHTLenS}: We consider the full likelihood of the weak gravitational lensing data from blue galaxy sample compiled in 
\cite{cg2}. \\

The base parameters set for the DM-DE scattering model, to be constrained, is given by
\begin{eqnarray*}
\label{P1}
P = \{100\omega_{\rm b}, \, \omega_{\rm dm},  \, 100\theta_{s}, \, \ln10^{10}A_{s}, \,  
n_s, \, \tau_{\rm reio},  \, \xi, \,  \, w_{\rm de} \},
\end{eqnarray*}
where the first six parameters are the base parameters for the standard $\Lambda$CDM model (see \cite{Planck2015} for more details) while the remaining two parameters correspond to its extension in the present study.
 
For DE properties, we assume that its equation of state parameter is constant ($w_{\rm de}= {\rm constant}$)
and we have taken $c^2_s=\delta p_{\rm de}/ \delta \rho_{\rm de} = 1$ ($c^2_s$ is in units of the speed of light). 
This necessarily implies that DE is a very light scalar field, as a typical canonical scalar field model with ${\rm mass} \lesssim H_0$. 
Also, in order to avoid the unphysical sound speed, usually we take $c^2_s = 1$. We consider suitable uniform priors on the parameters of the model 
under consideration. In particular, we choose $\xi \in [10^{-9}, 10^{-3}]$.  We take $10^{-3}$ as an upper bound, since values greater than that 
return surreal values to $\sigma_8$. On the other hand, it is reasonable to expect that $\xi << 1$.

\begin{table*}[hbt!]
\caption{\label{tab1} Constraints on the free parameters and some derived parameters of the DM-DE scattering model. 
Mean values of the parameters are displayed with 1$\sigma$ and 2$\sigma$ CL except the parameter $\xi$ for which the 2$\sigma$ upper bound is mentioned. The parameter $H_0$ is in the units of km
s${}^{-1}$ Mpc${}^{-1}$. Final row carries the $\chi^2_{min}/2$ values for the three cases of data-fitting. }
     \begin{center}
\begin{tabular} { ll  l l }
\hline
 Parameter  &CMB        & CMB + BAO & CMB +BAO +CFHTLenS \\
 \hline
 \hline
 
$100\omega_{\rm b }$& $2.227^{+0.014 +0.027}_{-0.014 -0.026}$& $2.231^{+0.014 +0.030}_{-0.016 -0.028}$  & $2.238^{+0.013 +0.026}_{-0.013 -0.025}$ \\                  
                                     
$\omega_{\rm dm }  $&$0.1193^{+0.0013 +0.0026 }_{-0.0015 -0.0024}$ & $0.1186^{+0.0013 +0.0024 }_{-0.0013 -0.0024}$ & $0.1179^{+0.0012 +0.0023}_{-0.0011 -0.0024}$  \\
                                   
$100\theta_{s }  $ &$1.04189^{+0.00029 +0.00056}_{-0.00029 -0.00057}$& $1.04195^{+0.00028 +0.00055}_{-0.00028 -0.00055}$ & $1.04191^{+0.00027 +0.00060}_{-0.00031 -0.00058}$\\

 $\ln10^{10}A_{s }$& $3.060^{+0.020 +0.043}_{-0.024 -0.041}$& $3.069^{+0.023 +0.048}_{-0.027 -0.043}$ & $3.061^{+0.022 +0.047}_{-0.027 -0.043}$ \\
                   
$n_{s }         $& $0.9649^{+0.0042 +0.0080}_{-0.0042 -0.0079}$& $0.9667^{+0.0048 +0.0095}_{-0.0048 -0.0096}$ & $0.9683^{+0.0045 +0.0091}_{-0.0045 -0.0086}$ \\
                                 
$\tau_{\rm reio }   $& $0.064^{+0.011 +0.022}_{-0.013 -0.022}$& $0.069^{+0.013 +0.027}_{-0.015 -0.025}$ & $0.066^{+0.012 +0.026}_{-0.015 -0.024} $ \\

$\xi               $&$<9.8\times 10^{-5}$& $<9.8\times 10^{-5}$ & $<9.4\times 10^{-5}$ \\
                        
$w_{\rm de}        $&$-1.11^{+0.14 +0.37}_{-0.25 -0.30}$ & $-1.06^{+0.07 +0.12}_{-0.06 -0.13}$ & $-1.03^{+0.06 +0.11}_{-0.06 -0.12}$ \\
          
\hline
$H_0            $&$71.2^{+7.7+10}_{-5.1 -10}$ &$69.6^{+1.5 +3.4}_{-1.8 -3.2}$ &$68.9^{+1.4 +3.0}_{-1.6 -2.9} $  \\

$\sigma_8        $& $0.848^{+0.067 +0.089}_{-0.044 -0.100}$& $0.834^{+0.018 +0.039}_{-0.018 -0.036}$ & $0.819^{+0.016 +0.032}_{-0.016 -0.030} $ \\

\hline

$\chi^2_{min}/2$ &  $6475.61$        &  $6481.87$     &   $6532.69$           \\
\hline
\end{tabular}
\end{center}
\label{tab1}
\end{table*}

\begin{figure}[h]
\includegraphics[width=8.5cm]{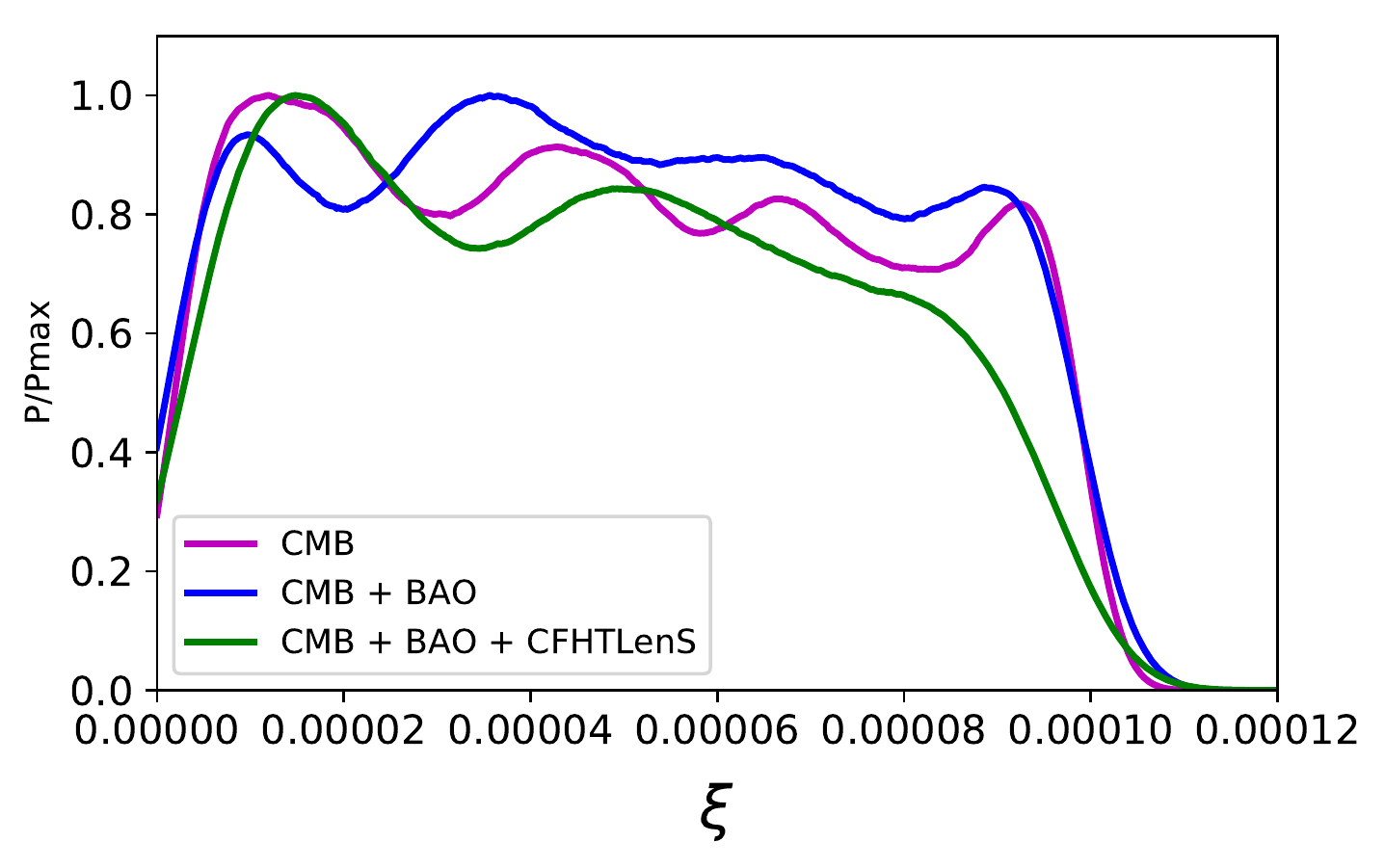}
\caption{\label{fig1} {\it{One-dimensional marginalized probability distribution of $\xi$}}}
\end{figure}

\begin{figure}[h]
\includegraphics[width=9cm]{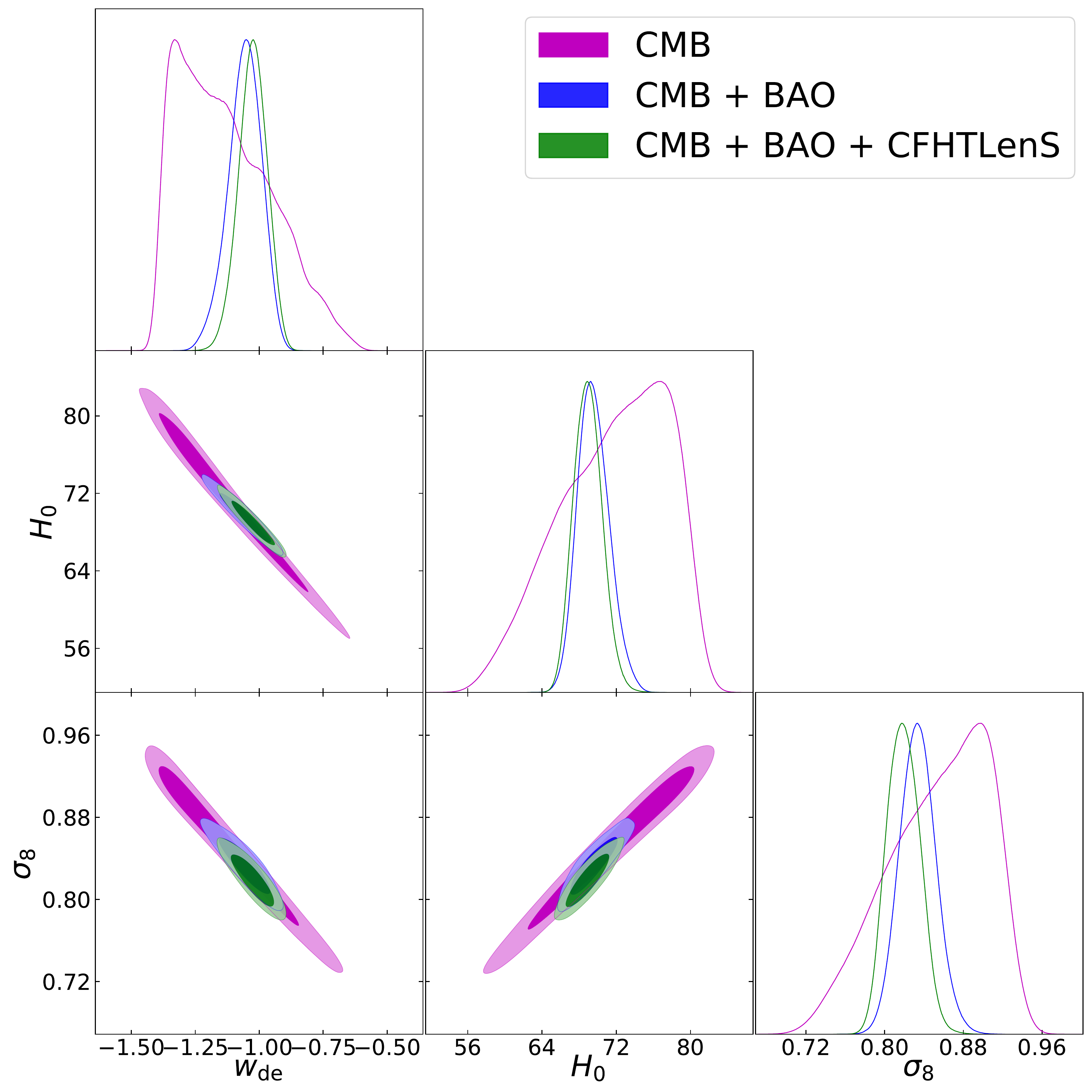}
\caption{\label{fig2} {\it{One-dimensional marginalized distribution,  two-dimensional 1$\sigma$ and 2$\sigma$  confidence contours for some selected parameters.}}}
\end{figure}

We modified the publicly available CLASS \cite{class} and Monte Python \cite{monte} codes for the DM-DE scattering model 
under consideration, and constrained the model parameters by utilizing three different combinations of data sets: CMB alone, CMB + BAO and
CMB + BAO + CFHTLenS. We used Metropolis Hastings algorithm on the model parameters 
to obtain correlated Markov Chain Monte Carlo (MCMC) samples from CLASS/Monte Python code, and finally analyzed these 
samples by using the GetDist Python package \cite{antonygetdist}.

Table \ref{tab1} summarizes the main results of the statistical analysis
carried out using three different combinations of data sets: CMB alone, CMB + BAO and
CMB + BAO + CFHTLenS. The one-dimensional marginalized distribution of $\xi$ is shown in Fig. \ref{fig1} while one-dimensional marginalized distribution, 
two-dimensional 1$\sigma$ and 2$\sigma$  confidence contours for some selected parameters are displayed in Fig. \ref{fig2}.

In all the three cases, we note that $\xi \lesssim 10^{-4}$. From eq. (\ref{xi}), we can write the cross section between DM and DE as
\begin{align}
  \sigma_{ d} < 10^{-29} \, {\rm cm^2} \, \Big( m_{\rm dm} \frac{c^2}{\rm Gev} \Big),
\end{align}
or in terms of the Thomson cross section value, $\sigma_T \simeq 10^{-25}$ cm$^2$, as

\begin{align}
  \sigma_d < 10^{-4} \sigma_T \, {\rm cm^2} \, \Big( m_{\rm dm} \frac{c^2}{\rm Gev} \Big).
\end{align}

The typical DM particle mass (as weakly interacting massive particle (WIMP) candidate) can be taken in the range 1 - 1000 ${\rm GeV}/c^2$.
Recently, the best upper limits for WIMP masses are found as 10 ${\rm GeV}/c^2$, 40 ${\rm GeV}/c^2$, and 50 ${\rm GeV}/c^2$
from the experiments XENON1T \cite{XENON1T}, PandaX-II \cite{PandaX}, and LUX \cite{lux}, respectively.
In order to qualitatively discuss our results, taking $m_{\rm dm} = 10$ ${\rm GeV}/c^2$, 
we expect to have $\sigma_d \lesssim 10^{-3} \sigma_T \, {\rm cm^2}$ at 95\% CL.
As expected, we can note that the interaction (non-gravitational) between the dark components is too small, at 
least three orders of magnitude lower than the photon-electron interaction.

The constraints obtained by using CMB data alone show a small preference for a phantom dynamics, i.e., $w_{\rm de} \lesssim -1$. Such a constraint on $w_{\rm de}$ from CMB has also been observed in \cite{w1, w2, w3}. 
In the present study, we may infer that the DM particles undergo elastic scattering
with scalar fields having negative kinetic term. It is well known that these fields suffer from instabilities at the classical 
and quantum levels \cite{w4,w5} that casts doubts about their existence. Nevertheless, 
observationally such fields cannot be discarded, as we have also observed here.

\begin{figure}
\includegraphics[width=8.5cm]{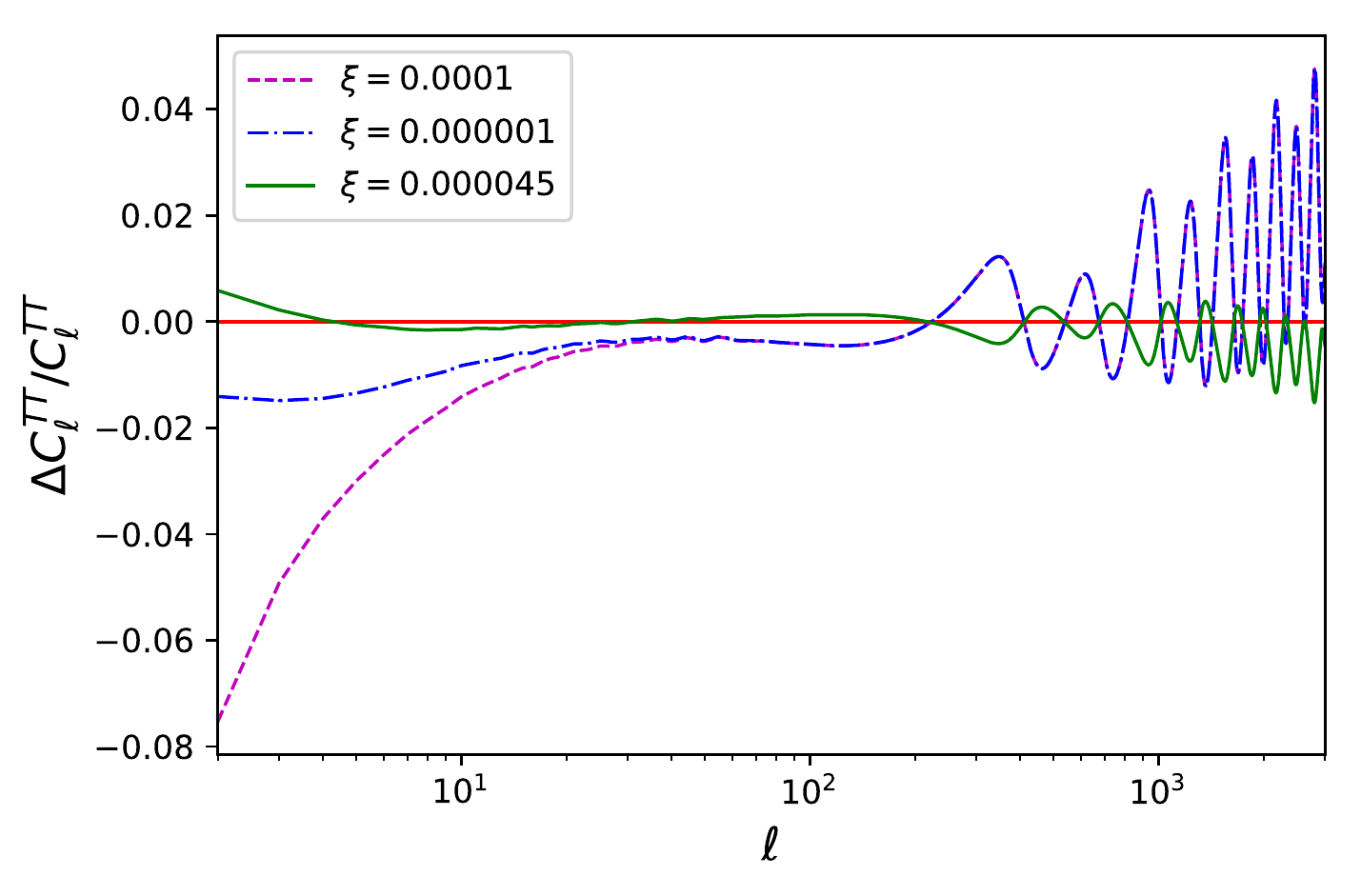}
\caption{\label{fig3} {\it{Relative deviation of CMB TT power spectrum from the base line Planck 2015 $\Lambda$CDM model (red line) for various values of $\xi$ while the other parameters are fixed to their bestfit mean values as given in Table \ref{tab1}.}}}
\end{figure}
\begin{figure}
\includegraphics[width=8.5cm]{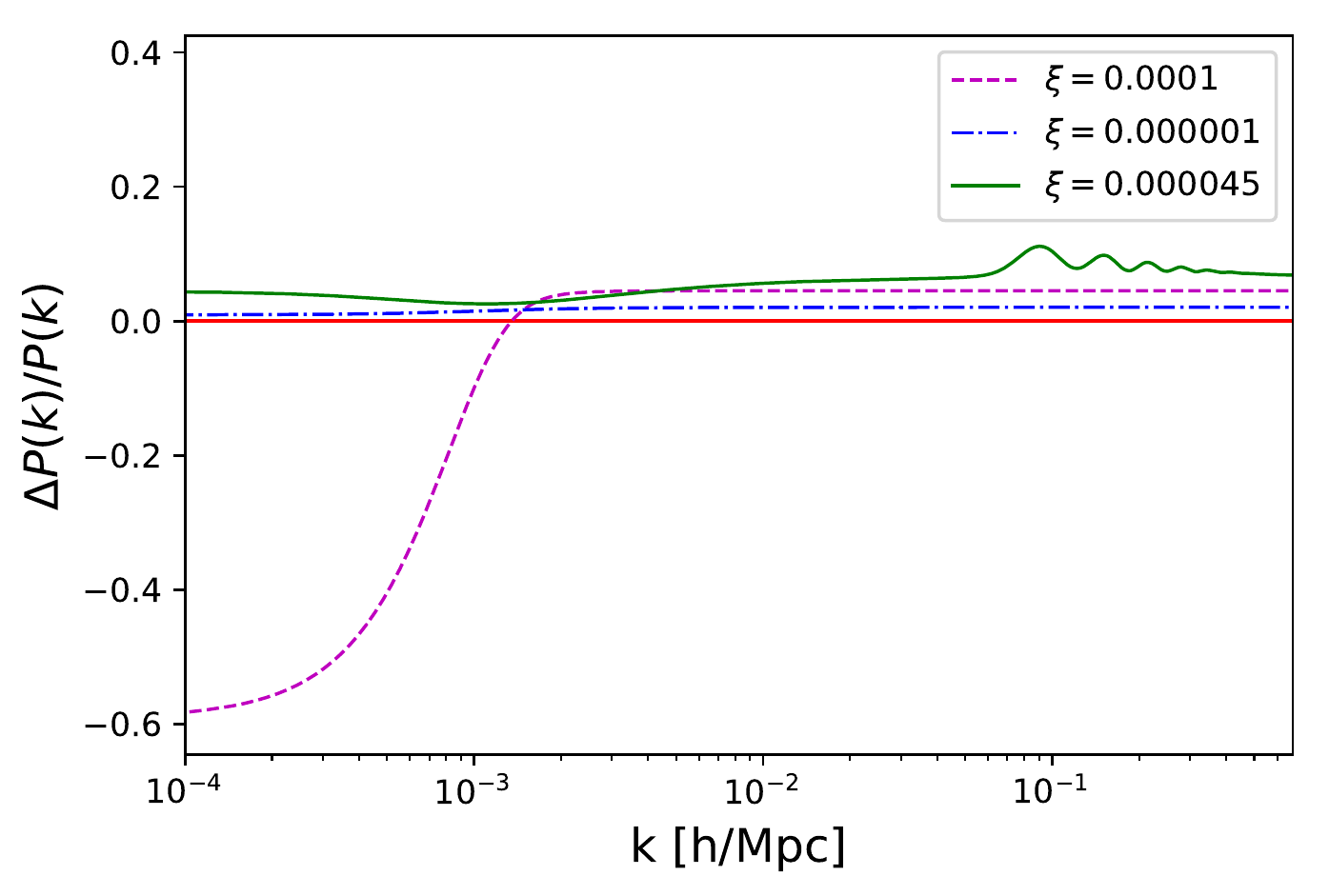}
\caption{\label{fig4} {\it{Relative deviation of matter power spectrum from the base line Planck 2015 $\Lambda$CDM model (red line) for various values of $\xi$ while the other parameters are fixed to their bestfit mean values as given in Table \ref{tab1}.}}}
\end{figure}

From Table \ref{tab1} and Fig. \ref{fig2}, it is clear that the combined data set CMB + BAO + CFHTLenS yields the most tight constraints on the model parameters. In order to observe/show the quantitative effects of $\xi$ on the power spectrum $P(k)$ of matter at $z=0$, and CMB TT power spectrum in contrast with the base line Planck 2015 $\Lambda$CDM model \cite{Planck2015}, we select some particular values of $\xi$ from its 95\% confidence region given by the combined data set CMB + BAO + CFHTLenS. 
Figures \ref{fig3} and \ref{fig4} depict the relative deviation of CMB TT and matter power spectra respectively, from the base line Planck 2015 $\Lambda$CDM model (red line) for the three values of $\xi$ as mentioned in the legends of the figures while the other parameters are fixed to their bestfit mean values as given in Table \ref{tab1}. We see that relative deviation of CMB TT power spectrum varies from the percent level to 8\%. On the other hand, significant relative deviation of matter power spectrum can be observed for the different values of $\xi$ in Fig. \ref{fig4}. It is clear that the larger values of $\xi$ tend to cause significant effects on the two power spectra.

\section{Conclusion and Perspectives}
\label{final}
In this study, we have considered an elastic scattering between DM and DE to constrain the cross section between the dark components.
This elastic scattering for DM-DE is analogous to the Thomson scattering for baryons and photons.
We find $\sigma_d \lesssim 10^{-29} \, {\rm cm^2}$ at 95\% CL from our joint analysis (CMB + BAO + CFHTLenS), for typical 
DM mass of the order 1-10 ${\rm GeV}/c^2$. This quantifies a very small interaction among the dark components. 
We find that the combined data set CMB + BAO + CFHTLenS puts the most tight constraints on the model parameters when compared with the other two cases considered in this study, and some possible values of $\xi$ can cause significant changes in the matter power spectrum $P(k)$. 

It may be noted that the DM-DE scattering model studied here reduces to the standard $\Lambda$CDM model with $\xi=0$ and $w_{\rm de}=-1$. The constraints in Table \ref{tab1} for the combined data set do not differ considerably from the ones  obtained in \cite{Planck2015} for the base line Planck 2015 $\Lambda$CDM model. Therefore, the presence of the non-zero parameter $\xi$ does not yield significant changes in the background cosmological dynamics. However, as pointed out in \cite{Simpson}, the elastic scattering among dark components may influence the growth of large scale structure in the Universe (see Figures \ref{fig3} and \ref{fig4}). The precise and robust constraints obtained in the present study may be utilized for studying the linear and nonlinear structure formation in the DM-DE elastic scattering model via N-body simulations (see \cite{Baldi1,Baldi2}). 
On the other hand, one can generalize the DM-DE elastic scattering scenario investigated here by considering an energy exchange between DM and DE, beyond
the pure momentum exchange, that is, a modified model with cosmological effects at background level (expansion history) 
and perturbative level (growth of structures) using a general parametrization of DE coupled to DM. 
Also, it could be worthwhile to investigate an elastic scattering between DE and massive neutrinos
(and/or dark radiation). Progress in this direction will be reported in a forthcoming paper.

\section*{Acknowledgments}
The authors are grateful to Bharat Ratra for constructive and fruitful discussions. S.K. gratefully acknowledges the support from SERB-DST project No. EMR/2016/000258.


\end{document}